# Model-based perfusion reconstruction with time separation technique in cone-beam CT dynamic liver perfusion imaging

Hana Haseljić[1,2] | Robert Frysch[1,2] | Vojtěch Kulvait[3] | Thomas Werncke[4,2] | Inga Brüsch[5] | Oliver Speck[2] | Jessica Schulz[6,2] | Michael Manhart[6] | Georg Rose[1,2]

[1]Institute for Medical Engineering, Otto von Guericke University Magdeburg, Magdeburg, Germany

[2]Research Campus STIMULATE, Otto von Guericke University Magdeburg, Magdeburg, Germany

[3]Institute of Materials Physics, Helmholtz-Zentrum Hereon, Geesthacht, Germany

[4]Institute of Diagnostic and Interventional Radiology, Hannover Medical School, Hannover, Germany

[5]Institute for Laboratory Animal Science, Hannover Medical School, Hannover, Germany

[6]Siemens Healthineers AG, Forchheim, Germany

**Correspondence**
Hana Haseljić, Institute for Medical Engineering, Otto von Guericke University Magdeburg, Magdeburg, Germany.
Email: hana.haseljic@ovgu.de

**Funding information**
Bundesministerium für Bildung und Forschung, Grant/Award Numbers: 13GW0473A, 13GW0473B

## Abstract

**Background:** The success of embolization, a minimally invasive treatment of liver cancer, could be evaluated in the operational room with cone-beam CT by acquiring a dynamic perfusion scan to inspect the contrast agent flow.

**Purpose:** The reconstruction algorithm must address the issues of low temporal sampling and higher noise levels inherent in cone-beam CT systems, compared to conventional CT.

**Methods:** Therefore, a model-based perfusion reconstruction based on the time separation technique (TST) was applied. TST uses basis functions to model time attenuation curves. These functions are either analytical or based on prior knowledge (PK), extracted using singular value decomposition of the classical CT perfusion data of animal subjects. To explore how well the PK can model perfusion dynamics and what the potential limitations are, the dynamic cone-beam CT (CBCT) perfusion scan was simulated from a dynamic CT perfusion scan under different noise levels. The TST method was compared to static reconstruction.

**Results:** It was demonstrated on this simulated dynamic CBCT perfusion scan that a set consisting of only four basis functions results in perfusion maps that preserve relevant information, denoise the data, and outperform static reconstruction under higher noise levels. TST with PK would not only outperform static reconstruction but also the TST with analytical basis functions. Furthermore, it has been shown that only eight CBCT rotations, unlike previously assumed ten, are sufficient to obtain the perfusion maps comparable to the reference CT perfusion maps. This contributes to saving dose and reconstruction time. The real dynamic CBCT perfusion scan, reconstructed under the same conditions as the simulated scan, shows potential for maintaining the accuracy of the perfusion maps. By visual inspection, the embolized region was matching to that in corresponding CT perfusion maps.

**Conclusions:** CBCT reconstruction of perfusion scan data using the TST method has shown promising potential, outperforming static reconstructions and potentially saving dose by reducing the necessary number of acquisition sweeps. Further analysis of a larger cohort of patient data is needed to draw final conclusions regarding the expected advantages of the TST.

## 1 | INTRODUCTION

Dynamic computed tomography perfusion (dCTp) imaging of the liver is an important non-invasive modality to inspect, quantify, and visualize haemodynamic changes in the liver. It is widely used to diagnose and characterize liver cancer, to plan and guide treatment, and to assess the response to treatment.[1,2] There are several treatment options for liver cancer, including locoregional therapies such as transarterial embolization (TAE) and transarterial chemoembolization (TACE). Recently, it has become increasingly common to perform so-called parenchymal perfusion imaging using cone-beam CT (CBCT)[3] to evaluate liver perfusion intra-procedurally.[4–9] For minimally invasive treatments, it would be highly beneficial to use CBCT, and related CBCT perfusion (dCBCTp) acquisition protocols, directly within the interventional suite, eliminating the need to move the patient outside of the operating room.[3,5] This approach not only allows the detection of tumor-feeding vessels but also enables real-time assessment of whether blood flow to the targeted region has been successfully blocked.[4]

The dynamic liver perfusion CBCT protocol used in this work was previously established,[10] which demonstrated the feasibility of dCBCTp imaging in animal models of liver perfusion by adapting existing protocol for brain perfusion.[11–13] In this protocol, a contrast agent is injected intravenously to enhance the visibility of the organ's internal structures. After the injection, a series of time-resolved CT volume images is acquired. For parenchymal perfusion imaging, only one rotation (sweep) around the patient is performed after the injection. In one sweep, lasting ∼ 4 s, 2D projections covering 200° angle are acquired. The volume is then reconstructed using all the 2D projections as if all were acquired at the same time point, disregarding the dynamic nature of perfusion.[14] For dynamic perfusion imaging first, more sweeps need to be acquired[10] and second, the reconstruction approach should take into account for the time dependency of every acquired 2D projection. Reconstructing the proposed ten separate sweeps[10] over a complete scan duration of ∼ 52 s and determining perfusion in this way would be subject to significant errors.[11]

A model-based approach can utilize the fact that data from every projection angle are recorded multiple times, that is, once in every sweep, to mitigate undersampling problems. In case of the dCBCTp perfusion of the brain, the so-called time separation technique (TST) was used to denoise the data and accurately estimate time attenuation profiles of the contrast agent.[15–17] The TST uses an orthogonal basis functions set (BFS) to model the perfusion dynamics, so the reconstruction problem is simplified and the overall computational time is reduced. In this research applicability and potential advantages of TST for liver perfusion imaging with dCBCTp were investigated. The algorithms were tested using animal models, where the dCTp and matching dCBCTp scans of an in vivo swine liver with embolized tissue were acquired.[18]

## 2 | METHODOLOGY

### 2.1 | Data acquisition

Three corresponding dCTp and dCBCTp scans of three different pig livers were acquired with a SOMATOM Force CT and an ARTIS pheno robotic C-arm system (Siemens Healthineers AG, Forchheim, Germany). The three animals were labeled with numbers from 1 to 3. The study was conducted in accordance with the European Directive 2010/63/EU and with the German law for animal protection (TierSchG). All experiments were approved by the local animal ethics committee (Lower Saxony State Office for Consumer Protection and Food Safety, LAVES 18/2809).

The three dCBCTp scans were acquired using the same experimental acquisition protocols[10] (see Table 1). The subsegment artery of the right hepatic artery was embolized with Onyx(R) (Medtronic, Meerbusch, Germany). The contrast material Imeron 300 was injected into the right hepatic artery. The C-arm performs 10 rotations in a bidirectional manner, pausing between

**TABLE 1** The acquisition protocol for dynamic CBCT perfusion scan.

| Duration | ∼ 52 s |
|---|---|
| Number of 3D rotations | 10 |
| Number of views per rotation | 248 |
| Angle covered | 200° |
| Angle step | 0.8° |
| Detector size (no. of pixels) | 624 × 464 |
| Pixel spacing | 0.16 × 0.16 mm$^2$ |
| Tube peak voltage | 90 kV |

Abbreviation: CBCT, cone-beam CT.

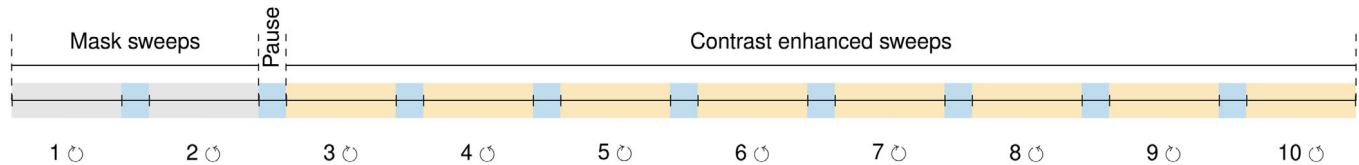

**FIGURE 1** Schema of dynamic CBCT perfusion scan with ten sweep protocol. CBCT, cone-beam CT.

consecutive rotations. In this way, five forward and five backward sweeps are acquired. The first two sweeps are mask sweeps, one for the forward and one for the backward rotation as shown in Figure 1. Each sweep comprises 248 views. There is an inconsistency between the angles at which the corresponding views with index $i$ in forward and index $(248 - i)$ in backward sweep are acquired. The time intervals between two consecutive views, that is, frame time, differ within the sweep due to the accelerating rotation of the C-arm at the beginning of the scan and decelerating towards the end of the scan. The duration of all dCBCTp scans, including all 10 rotations, is $\sim 52$ s. The rotation time lies around 3.9 s with the pause time between two rotations of $\sim 1.4$ s.

The scanning durations of the dCTp scan for animals 1, 2, and 3, were 65.997 s, 56.997 s and 41.998 s, respectively. The tube peak voltage for the dCTp scans was set to 90 kv. The matching dCBCTp and dCTp scans have the same contrast agent injection protocols as documented in Table 2. The animals were in general anesthesia with muscle relaxation to avoid any muscle movements such as breathing.

## 2.2 | CBCT scan simulation

As the different positioning of the animal in CT and in the corresponding CBCT makes it difficult to compare both volumes voxel-wise, the dCBCTp scan was simulated by reprojecting the dCTp scan(rdCBCTp). Due to data limitation (see Section 2.4.2) only the shortest dCTp scan was reprojected. To maintain a realistic rotation velocity of the C-arm, the rdCBCTp scan consists of eight contrast-enhanced sweeps. The time-resolved volumes are interpolated using Akima splines,[19] and resampled at time points defined by the frame times of the original corresponding dCBCTp scan during eight sweeps. These were estimated from the DICOM header

**TABLE 2** Contrast agent injection protocols for three scans.

| | Contrast agent | | | |
|---|---|---|---|---|
| Scan | Dose (mL) | Flow rate (mL/s) | Flow duration (s) | Volume (mL) |
| 1 | 14.0 | 3.0 | 7.0 | 20.0 |
| 2 | 10.43 | 2.8 | 5.07 | 14.9 |
| 3 | 10.43 | 2.9 | 5.1 | 14.9 |

*Frame Time Vector*. The forward projector algorithm of the CT Library[20] was used to simulate dCBCTp projection data from the dCTp volumes, where the projection matrix of the forward sweep of the real dCBCTp scan was used to model the geometry. To avoid the need for 2D-2D registration to compensate for angle inconsistency of forward and backward sweeps, the order of the forward projection matrix was reversed to estimate the backward sweep geometry. Although the breathing was suppressed, some motion is still present in the intestines.

The voxel size for reconstruction is the same as the voxel size of the matching CT scan, (0.7305 mm, 0.7305 mm, and 1.5 mm). The voxel count in one volume is $512 \times 512 \times 175$.

### 2.2.1 | Noise addition

Considering that the CBCT scans suffer from higher noise levels than CT, two different levels of Poisson noise, higher noise of $2.1 \times 10^5$ photons per mm$^2$ and moderate noise of $6 \times 10^5$ photons per mm$^2$ at the detector, were added to reprojected dCBCTp scan. The noise level was determined by adjusting the standard deviation in scan of brain phantom to the standard deviation estimated in scan of water cylinder phantom.[21,22] The effect of the noise added on the reconstruction images is shown in Figure 2.

## 2.3 | Reconstruction

The reconstruction from the dCTp scan is a 4D reconstruction task because the contrast enhancement varies dynamically during the acquisition of the volume time series. Each voxel in the volume is defined by the four coordinates, $(x, y, z, t)$, where $x, y, z$ are spatial variables and $t \in \mathcal{I}$ represents a time point during the scan duration $\mathcal{I}$. The time attenuation curve (TAC) describes the contrast agent dynamics in each voxel at all sampling time points $t$. TACs are extracted from the reconstructed volumes. Therefore, the reconstruction problem is a time-dependent CT problem

$$\mathbf{A}\mathbf{v}(t) = \mathbf{p}(t) \quad t \in \mathcal{I} \qquad (1)$$

where $\mathbf{A}$ is the system matrix that maps volume $\mathbf{v}$ at time point $t$ onto the projection space $\mathbf{p}$.

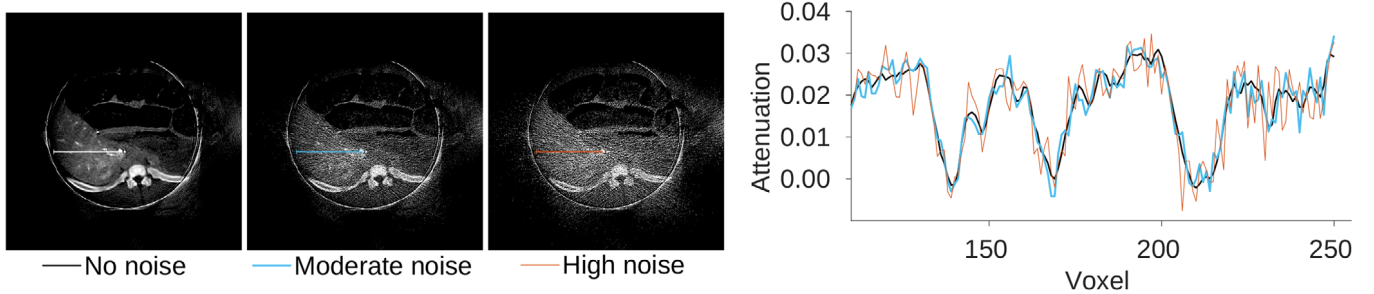

**FIGURE 2** Effect of moderate and high noise level to reconstructions.

In a dCBCTp scan at a single time point $t \in [0, T_s]$, where $T_s = 4s$ is the duration of one sweep, a 2D projection for one source-detector position, that is, angle $\varphi_i, i \in \{1, .., \Phi\}$, is acquired. In ten-sweep acquisition protocol, only 10 projections for the same $\varphi_i$ are acquired over the scan duration $T = 52s$. This makes the reconstruction problem Equation 1 in the domain of dCBCTp scan reconstruction underdetermined. The time interval for the same $\varphi_i$ between every two sweeps, due to a bidirectional rotation of the C-arm and pause time between sweeps is not constant.

In the straightforward reconstruction approach,[17] each sweep is reconstructed separately under the assumption that all the projections in one sweep have been acquired at the same time point, and thus the dynamic change of contrast agent flow through the organ is not properly represented. This limits the quantifiability of the perfusion maps.

The TST[16,17] relies on the idea of model-based reconstruction (MBR). Both voxels' dynamics $\mathbf{v}(t)$ and projections $\mathbf{p}(t)$ are modeled as the linear combination of the set of orthonormal functions

$$\mathcal{B} = \{\Psi_1, \ldots, \Psi_N\}, \quad N \to \infty. \quad (2)$$

Furthermore, the TST assumes that with a very suitable orthonormal functions (see Section 2.4.2), only the first $\hat{N} \leq 5$ elements would be sufficient to model perfusion. In this paper the set $\mathcal{B}$ is called a basis and its functions are basis functions.

Each voxel $v_\nu(t)$ is represented as

$$v_\nu(t) = \sum_{i=1}^{\infty} w_{\nu,i} \Psi_i(t) \approx \sum_{i=1}^{\hat{N}} w_{\nu,i} \Psi_i(t) \quad (3)$$

and each projection pixel $p_n$

$$p_n(t) = \sum_{j=1}^{\infty} c_{n,j} \Psi_j(t) \approx \sum_{j=1}^{\hat{N}} c_{n,j} \Psi_j(t), \quad (4)$$

where $\nu$ codes $(x, y, z)$ coordinates in the volume and $n$ codes $(l, k, \varphi)$ coordinates in the projection images.

Using this notation, the reconstruction problem in Equation 1 takes the form

$$\mathbf{A} \sum_{i=1}^{\hat{N}} \mathbf{w}_i \Psi_i(t) = \sum_{j=1}^{\hat{N}} \mathbf{c}_j \Psi_j(t) \quad t \in \mathcal{I}. \quad (5)$$

As the basis functions form a set of orthogonal vectors, the scalar product of two differing basis functions is zero. By performing the scalar product on both sides of Equation 5

$$\mathbf{A} \sum_{i=1}^{\hat{N}} \mathbf{w}_i \langle \Psi_i(t), \Psi_l(t) \rangle = \sum_{j=1}^{\hat{N}} \mathbf{c}_j \langle \Psi_j(t), \Psi_l(t) \rangle \quad t \in \mathcal{I}, \quad (6)$$

the reconstruction problem Equation 1 becomes $\hat{N}$ separate CT problems

$$\mathbf{A}\mathbf{w}_i = \mathbf{c}_i \quad i \in \hat{N}, \quad (7)$$

which simplifies the reconstruction task in such a way that the number of reconstructions that need to be performed equals the number of basis functions $\hat{N}$ used to model the time dynamics.

Now, $\mathbf{c}$ from Equation 4 are determined by calculating the scalar product

$$\langle \mathbf{p}(t), \Psi(t) \rangle. \quad (8)$$

The calculation of the scalar product is problematic since it assumes that the basis functions are already sampled at the time points at which the projections are acquired. To carry out the integration for the scalar product estimation, the interpolation of the projections would be acquired, which would lead to significant errors. To circumvent this problem, an optimization approach was applied as follows:

$$\min_{\{c\}} \|\Psi(t) \cdot \mathbf{c} - \mathbf{p}(t)\|^2 =: \mathbf{c}'. \quad (9)$$

Instead of determining the coefficients $\mathbf{c}$ from the scalar products, they are defined as the parameters that best approximate the function $p$. It should be noted that this fitting method leads to the following approximation

$$p(t) = \sum_{i=1}^{\infty} c_i \Psi_i(t) \approx \sum_{i=1}^{\hat{N}} c'_i \Psi_i(t). \quad (10)$$

The prerequisite step is to resample the basis functions at time points specific to each pixel $p_{(x,y,\varphi)}$. Subsequently, these fitted projections are reconstructed and the coefficients $w_i, i \in \{1, .., \hat{N}\}$ are obtained. From these reconstructed coefficients, the time series of volumes for a given sampling can be computed.

The choice of a proper reconstruction algorithm is only limited by Equation 7. Here, all the reconstructions are computed using a Krylov method based iterative reconstruction to reduce the computation time.[23]

## 2.4 | BFS

The basis functions have to be suitable for modeling the contrast agent flow, that is, to approximate the contrast agent flow with as few as $\hat{N}$ basis functions. Thus, two different BFSs are used, an analytical BFS and a prior knowledge (PK) BFS.

### 2.4.1 | Analytical set

A set of trigonometric functions (see Equation 11) was successfully used to model brain perfusion.[17] Furthermore, the same BSF was also used to model liver perfusion using a real dCBCTp scan.[18] The first basis is a constant necessary to model static behaviour of the anatomical background.

$$\Psi_0 = 1, \Psi_1 = \sin\left(\frac{2\pi t}{T}\right), \Psi_2 = \cos\left(\frac{2\pi t}{T}\right), \Psi_3$$
$$= \sin\left(\frac{4\pi t}{T}\right), \Psi_4 = \cos\left(\frac{4\pi t}{T}\right) \quad (11)$$

### 2.4.2 | Prior knowledge set

The PK BFS extracted from the CT perfusion time-resolved volumes should optimally model the dynamics of liver perfusion.[24] Ideally, one generalized BFS could be selected which would be suitable for future applications on any subject for which the prior CTp scan was not available. Therefore, here, the PK BFS was extracted from CT time series of only two animals; see Figure 3. This BFS was tested against the third dCBCTp and rdCBCTp scan.

The organ was manually segmented in all slices of CT volumes. All the abdominal organs (intestines, gallbladder, stomach), bones (spine and ribs), surrounding vessels (e.g., gastric artery), and reconstruction arte-facts caused by embolization material were excluded. Orthonormal basis functions were extracted from these volumes using singular value decomposition (SVD).[25] The initial number of the singular vectors forming the set is determined by the elbow method.[26] To further narrow the number of basis functions forming the PK BFS, the perfusion maps, obtained with BFS consisting of different number of basis functions, were calculated and the Pearson correlation coefficients were calculated to compare them with CT perfusion maps.[24] The selection of basis functions is explained in more details in Appendix. Then, all remaining vectors were considered as noise and omitted. In this work the number of selected basis functions was $\hat{N} = 4$.

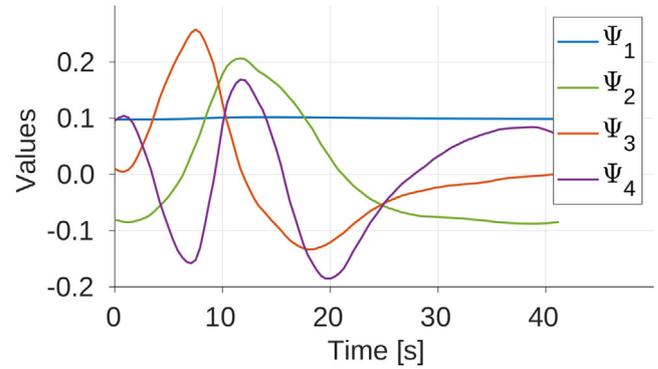

**FIGURE 3** PK BSF extracted from two animals. BSF, basis functions set; PK, prior knowledge.

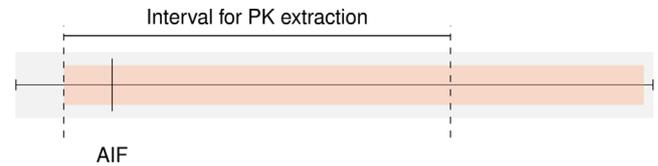

**FIGURE 4** Time alignment of two CT perfusion scans (gray and orange) with respect to the AIF for extraction of PK. AIF, arterial input function; PK, prior knowledge.

The SVD is applied only within the duration of the shortest scan — ∼ 42 s, meaning it is covering all eight sweeps in rdCBCTp scan. Applying SVD over longer scan duration, would assume that the volumes of the shortest scan wouldn't be able to cover the whole interval and the cross-validation would not be possible. For extraction of PK, the two scans were aligned in time with respect to the peak of the arterial input function(AIF). The aligning is performed by shifting one of the scans in time. The AIF was selected based on its peak value and time point at which this peak occurred. Figure 4. shows how the time alignment is performed w.r.t. the duration of the scans via the AIF.

### 2.4.3 | Time shifting of basis functions

The PK and the projections are time dependent and to model the contrast agent flow through the organ correctly, the PK might need to be shifted in time. The shift is determined after the time alignment of CBCT scan and basis functions interval. The time alignment is performed in the same manner as the alignment of the CT scans for the extraction of the PK. Two types of time shifting are to be differentiated but not mutually exclusive. The first type of the shifting occurs when the basis functions cannot support the whole CBCT scan interval. Then, all the values of the pixel $p_{(x,y,\varphi)}$ that were acquired outside the supported interval, at the beginning or at the end of the scan, will be excluded from the fit (see Figure 5). The second shift occurs when the basis functions interval has to be cropped. Then both, offset at the beginning and the offset at the end of the basis functions interval could be needed (see Figure 6).

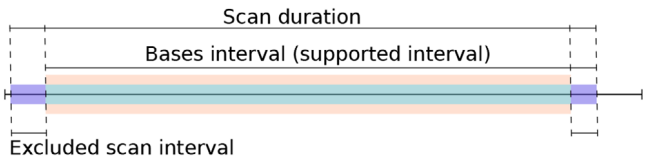

**FIGURE 5** Basis functions time shifting. Exclusion of projections (purple) from the CBCT scan which are not supported by the basis functions. CBCT, cone-beam CT.

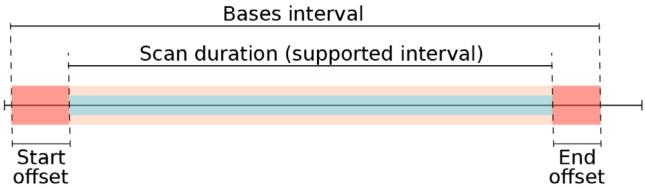

**FIGURE 6** Basis functions offsetting (red) when the covering interval is longer than the duration of the CBCT scan. CBCT, cone-beam CT.

For the data used in this work, only the first type of basis shifting was needed. For the real dCBCTp scan the projections of the first two mask sweeps were outside the supported interval, meaning that the fitting started at the beginning of the third sweep. The AIF of the rdCBCTp reaches its peak value 1.5 s after the AIF based on which the CT scans for the PK were aligned so all the projections from the first 1.5 s were excluded from the fitting.

## 2.5 | Calculation of perfusion maps

The final step in perfusion imaging is the calculation of the perfusion maps. Perfusion maps are the visualization of the perfusion parameters. The first volume is considered a mask sweep and is subtracted from all the following volumes in time series so that the perfusion parameters can be calculated.[27] With the TST, the value of each voxel in time point over the scanning interval is estimated using reconstructed coefficients and the BFS. The first estimated value is subtracted from the rest, so that the time attenuation curve represents only the time dynamics of the voxel.

Since a small amount of contrast agent was injected into the right hepatic artery at a high rate, no contrast agent is expected in the portal vein during the acquisition time. Therefore, only four perfusion parameters are calculated in the arterial phase; blood flow (BF), blood volume (BV), mean transit time (MTT), and time to peak (TTP). Previously, it was shown that the embolized regions are better distinguishable in perfusion maps calculated using the deconvolution method in comparison to the maximum slope approach. Also, the MTT can be miscalculated with the maximum slope approach even with the high injection rate of contrast agent.[28,29] Thus, the perfusion parameters are calculated here using a deconvolution algorithm.[30]

The perfusion parameters are calculated voxel-wise. The TAC of each voxel $v$, $tac_v(t)$ is represented as a convolution of AIF $aif(t)$ and convolution kernel $k_v(t)$

$$tac_i(t) = aif(t) * k_i(t). \quad (12)$$

This problem can be represented algebraically

$$\mathbf{Ak} = \mathbf{c} \quad (13)$$

where $\mathbf{A}$ is the Toeplitz matrix constructed from discretized AIF. Now, $\mathbf{k}$ is found by solving

$$\mathbf{k} = \mathbf{A}^{-1}\mathbf{c}. \quad (14)$$

SVD is applied on $\mathbf{A}$, then all singular values larger than the Tikhonov regularizer $\lambda_{rel} = 0.3 * \sigma_{max}$ are inverted, $\sigma = 1/\sigma$, and all smaller than $\lambda_{rel}$ set to $\sigma = 0$. This way the new matrix $\mathbf{A}'$ is formed and, $\mathbf{A}^{-1}$ is replaced by it, so that

$$\mathbf{k}' = \mathbf{A}'\mathbf{c} \quad (15)$$

can be estimated.[30] Once $\mathbf{k}'$ is estimated, the perfusion parameters for each voxel $v$ with TAC sampled in $n = 100$ points are calculated as follows:

$$BF_v = max\{k_{v,i}, i \in \{1..n\}\}, \quad BV_v = \sum_{i=1}^{n} k_v, \quad MTT_v = \frac{BV_v}{BF_v}. \quad (16)$$

The TTP is not dependent on the convolution kernel, but only on the TAC and it equals the time point at which the TAC reaches its peak enhancement.

For the visualization of perfusion maps, the ASIST[31] colormap is used. With this colormap, the hypo-perfused — in this case embolized (dark blue), region is easily distinguishable from the healthy perfused tissue. For all perfusion maps, the lower values are represented with cold colors and vice versa for higher values. A Gaussian blur with $\sigma = 3.0$ px is applied slice-wise to the organ in the perfusion map to ensure noise reduction and for features to be better distinguishable.

## 2.5.1 | Selection of AIF

The wrong selection of the AIF can result in a miscalculation of the perfusion parameters. The AIF can be selected in one of the contrast enhanced reconstructed volumes or in case of TST in reconstructed coefficient which corresponds to one of the basis functions modeling the contrast agent dynamics. The AIF was localized in a single voxel.[17] First, the region of interest (ROI) in a vessel in which the contrast agent was injected was selected. Here, the ROI was selected in the right hepatic artery. From this region a single voxel is selected to be the AIF. The AIF should be selected in the same region for CT and rCBCT,[32] and when possible in the real CBCT volume as well. Selecting AIFs in different vessels would not only affect the calculation of the perfusion parameter values, but also the overall quality of the perfusion maps. The AIF with the highest peak reached in the earliest time point was selected as the single-point AIF.

## 2.6 | Evaluation of perfusion maps

### 2.6.1 | Simulated dynamic CBCT perfusion scan

Simulation of the dCBCTp scan enables voxel-wise comparison of the perfusion maps by calculating the Pearson correlation for every slice. This is possible since the organ is not displaced compared to the dCTp scan. As a consequence, it has suffered from organ truncation not only in the *z*-axis but also in the *xy*-plane. In addition, the motion of the intestines was captured, which can result in streak artifacts. Thus, to analyze only reconstructed voxels that are not severely affected by these occurrences, an additional mask excluding these regions was used for calculation of Pearson correlation coefficients. To analyze the TST denoising potential, the standard deviation has been estimated in all perfusion maps for straightforward reconstruction, TST with analytical BFS and TST with PK BFS. The standard deviation has been estimated using SKIMAGE.RESTORATION.ESTIMATE_SIGMA function within the SCIKIT-IMAGE Python package.[33,34]

### 2.6.2 | Real dynamic CBCT perfusion scan

The position of the animal during the dCTp scan was not the same as during the real dCBCTp scan. To fit it into the field of view of the C-arm CT, the animal was shifted and tilted. Thus, only a qualitative assessment was performed through visual inspection. To find the corresponding slices, features in the slice, such as position of the gallbladder, stomach, ribs and spine were located and compared. Additionally, the standard deviation has been estimated as explained in Section 2.6.1.

## 3 | RESULTS

### 3.1 | Simulated dynamic CBCT perfusion scan

In Figures 7, 8, and 9, the AIFs are shown for all three noise level scenarios in rCBCT, no noise, moderate noise, and high noise level, after the additional Gaussian smoothing was added to CT and CBCT volumes for calculation of perfusion parameters. The TAC for AIF in CT is obtained for the denoted voxels from all time series volumes, in the static reconstruction from reconstructed volumes for every sweep, and in MBR with the TST - from time-resolved volumes. The noise affects the AIF reached peak values but not the overall relation between them compared to the CT AIF. For all rCBCT AIFs, the peak value is lower compared to the reference CT AIF except for the TST with PK with high noise. The AIF peaks of TST with analytical and PK BFS are the closest in time to the reference CT AIF occurring. The same is observable in all three noise levels. Modeling the AIF with analytical basis functions (rCBCT TST-Analytical) can result in an additional peak so that it resembles the peak due to reperfusion or reinjection of contrast agent material. The occurrence of the AIF peak for static reconstruction almost 2 s later is due to an assumption that all the views within one sweep are considered to be acquired within the same time point. The AIF similarity in shape is observable in all noise scenarios for the AIF modeled by the PK (rCBCT TST-PK).

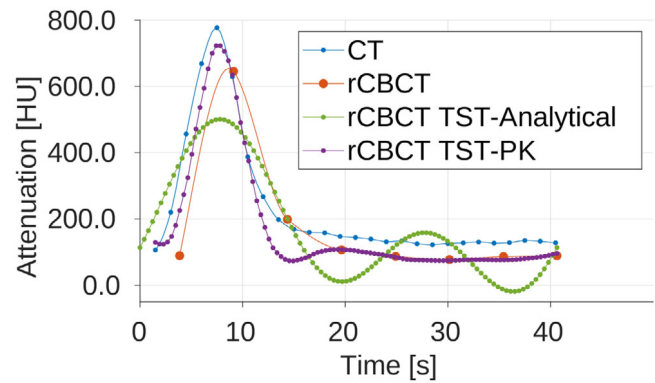

**FIGURE 7** AIF in reprojected dCBCTp scan with no noise added. AIF, arterial input function; dCBTp; cone-beam CT perfusion.

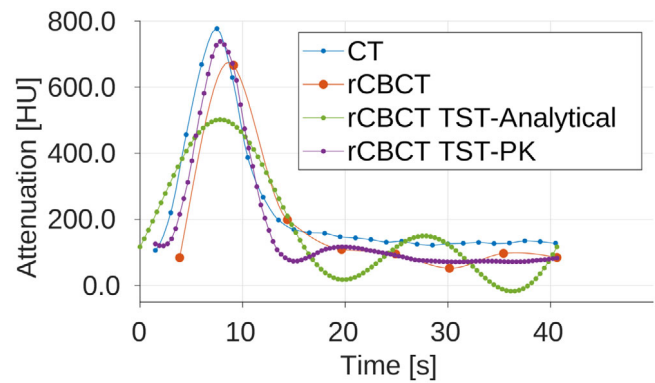

**FIGURE 8** AIF in reprojected dCBCTp scan with moderate noise level. AIF, arterial input function; dCBCTp; cone-beam CT perfusion.

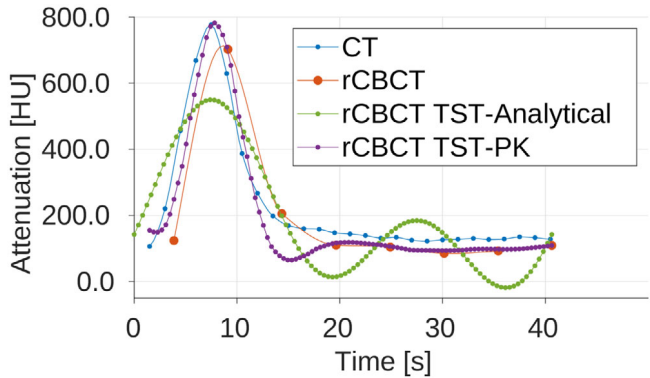

**FIGURE 9** AIF in reprojected dCBCTp scan with high noise level. AIF, Arterial input functions; dCBCTp; cone-beam CT perfusion.

The Pearson correlation coefficients between the CT perfusion maps and rCBCT perfusion maps using AIFs from the Figures 7, 8, and 9, for static reconstruction ("Static") and MBR by the means of TST using analytical ("Analytical") and PK basis functions ("Prior") are given in Table 3. The regions in slices with reconstruction artifacts were not taken into account. For the BF and MTT, the TST with PK basis functions outperforms static reconstruction for all three noise scenarios. As noise levels rise, the correlation values for static reconstruction are decrease. Even though the same behaviour is expected, and it is, for all three CBCT reconstructions, it is the most pronounced for the static reconstruction. For the MTT, the Pearson correlation is below 0.5 for static reconstruction and for TST with analytical basis functions, unlike 0.7118 for TST with PK for high noise. With the TST the change in correlation between different noise levels is smaller than in case of the static reconstruction, with BF remaining the most stable with PK. The change in AIF values and shape did not indicate these occurrences, since the shape of the CBCT AIFs for moderate and high noise is more similar to CT AIF, see Figures 8 and 9. For BV, the TST with analytical basis functions is the best and remains this way independently of the noise levels. However, the TST, neither with analytical nor PK basis functions, manages to outperform the static reconstruction for TTP, except for the high noise, for the insignificant difference of 0.0009.

The perfusion maps, BF, BV, MTT and TTP, for the slice at the middle of the liver are shown in Figure 11 for the scenario of moderate noise level. In the first column are the CT perfusion maps, then perfusion maps calculated for simulated dCBCTp scan in order as follows: static reconstruction ("Static"), TST using analytical BFS ("TST Analytical"), and TST using PK consisting of four basis functions ("TST Prior"). The hypo-perfused region, that is, embolized region, is easily distinguishable in all perfusion maps. The healthy tissue, that is, well perfused regions, are clearly visible as well. The misestimation of perfusion values due to streak artifacts are visible in the upper part of the organ for all four parameters in the rCBCT perfusion maps. For the BF, this is the most pronounced for TST with PK basis functions. In static reconstruction and in TST with PK basis functions, the underestimation in comparison to CT can be observed, particularly around the vessel structures. This region is pointed by white arrow in Figure 11. Nevertheless, the BV of TST with PK is the most similar to CT. For the MTT, the misestimation due to the mentioned reconstruction artifacts is observable in static reconstruction and for TST with PK basis functions. This region for the MTT in static reconstruction is indicated by a yellow arrow. For all four parameters, the most similar to CT perfusion maps are the ones depicted in the fourth column of the TST with PK. Under moderate noise, the TST was not yet able to outperform the static reconstruction for the TTP. This can be especially visible for the TTP with TST with PK. With increasing noise, the TST with PK appears to be more comparable to CT perfusion maps (see Table 3). The noise was reconstructed and is visible in perfusion maps, see second and third column, unlike for the TST with PK. Also, some overestimation is visible in static reconstruction, particularly for BV. The estimator of the standard deviation for TST gave consistently lower values than for the static reconstruction, see Table 4. The denoising is especially observable for BF and so with high noise level.

**TABLE 3** Pearson correlation coefficients for rCBCT perfusion maps compared to CT perfusion maps calculated. The highest values are highlighted in bold.

|     | No noise |        |        | Moderate noise |        |        | High noise |        |        |
|-----|----------|--------|--------|----------------|--------|--------|------------|--------|--------|
|     | Static   | TST    |        | Static         | TST    |        | Static     | TST    |        |
|     |          | Analytical | Prior |             | Analytical | Prior |          | Analytical | Prior |
| BF  | 0.9298   | 0.9112 | **0.9371** | 0.9135     | 0.8902 | **0.9259** | 0.8778 | 0.8597 | **0.9008** |
| BV  | 0.8710   | **0.8930** | 0.8475 | 0.8153     | **0.8451** | 0.8106 | 0.7379 | **0.7792** | 0.7576 |
| MTT | 0.7385   | 0.7051 | **0.8220** | 0.6341     | 0.5701 | **0.7778** | 0.4947 | 0.4225 | **0.7118** |
| TTP | **0.7735** | 0.7575 | 0.7620 | **0.7145**   | 0.7002 | 0.7065 | 0.6356 | 0.6281 | **0.6365** |

Abbreviations: BF, blood flow; BV, blood volume; MTT, mean transit time; TST, time separation technique; TTP, time to peak.

**TABLE 4** Standard deviation of the noise estimated in perfusion maps for reprojected CBCTp scan. The lowest values are highlighted in bold.

|     | Moderate noise | | | High noise | | | Real scan | | |
|-----|---|---|---|---|---|---|---|---|---|
|     | Static | TST | | Static | TST | | Static | TST | |
|     |        | Analytical | Prior |    | Analytical | Prior |    | Analytical | Prior |
| BF  | 0.706 | **0.520** | 0.582 | 1.420 | **0.910** | 1.057 | 0.324 | **0.149** | 0.243 |
| BV  | 0.071 | 0.061 | **0.042** | 0.144 | 0.109 | **0.076** | 0.026 | 0.014 | **0.009** |
| MTT | 0.105 | 0.098 | **0.091** | 0.177 | 0.157 | 0.151 | 0.056 | **0.026** | 0.041 |
| TTP | 0.299 | 0.251 | **0.239** | 0.534 | 0.355 | **0.313** | 0.157 | **0.129** | 0.145 |

Abbreviations: BF, blood flow; BV, blood volume; CBCTp, MTT, mean transit time; TST, time separation technique; TTP, time to peak; TST, time separation technique.

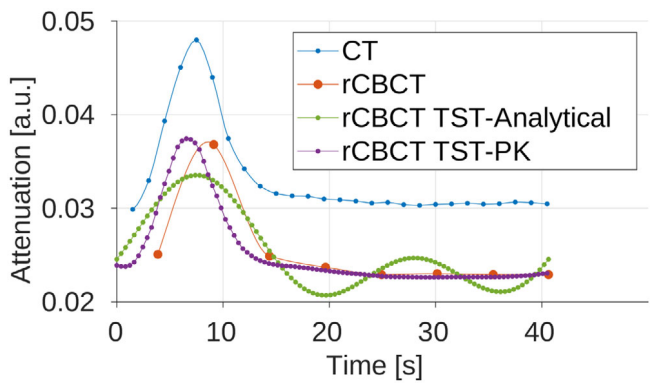

**FIGURE 10** AIF in real dCBCTp scan. AIF, Arterial input functions; dCBCTp; cone-beam CT perfusion.

## 3.2 | Real dynamic CBCT perfusion scan

In Figure 10 the AIFs of real dCBCTp scan of the same animal for which the CT scan was reprojected - in this work labeled 3, are shown in order as follows: CT, static reconstruction ("CBCT Static"), TST using analytical basis functions ("CBCT TST-Analytical") and TST using PK basis functions ("CBCT TST-PK"). All of the AIFs reach their peak at different time points, with the AIF modeled with TST using PK having the highest peak value. Note that due to different CT modalities, the values of perfusion parameters cannot be compared. The AIF for the static reconstruction is shifted the most, which is expected considering the time considerations for the acquired projections. The analytical basis functions do have the peak closest in time to the CT peak, but lower. The same was noted for the reprojected CBCT scan. The AIF with TST with PK is the closest to the CT AIF in peak value.

The perfusion maps for the dCBCTp scan of animal 3 is shown in Figure 12. The hypo-perfused region is well distinguishable in static and TST with analytical basis functions. However, for the PK, some overestimation can be noticed in some regions, which also makes the hypo-perfused region less pronounced - pointed by the yellow arrow. What additionally reduces the quality of perfusion maps are the artifacts, which result in streaks of overperfused regions that are not comparable to the CT scan. However, the MTT with analytical basis functions is the least comparable to CT perfusion maps. This is particularly visible in the left part of the organ. The BV of TST with analytical basis functions is showing some underestimation in the upper regions, but with the PK these overestimations still result in a perfusion map in general more similar to CT. For the MTT, the TST with PK is overall the most similar. However, the hypo-perfused region is still misinterpreted in terms of TTP - pointed by the white arrow. Similar to the simulated CBCTp scan, the TST results in noise reduction. However, for real CBCT, with analytical BFS, the estimated standard deviation is overall lower than for the PK BFS, see Table 4. The noise reduction is especially visible in BF.

## 4 | DISCUSSION

In this work, it was shown that the dynamic perfusion CBCT scan of eight sweeps is sufficient for dynamic liver perfusion imaging, which can be beneficial for minimally invasive treatments of liver cancers. Independently of the BFS used, the model-based perfusion reconstruction by the means of TST has provided the perfusion maps in which there is a clear distinction of hypo-perfused region, that is, the embolized region. Omitting last two sweeps, this scenario first leads to a shorter scan time and saving radiation dose compared to the proposed scanning protocol[10] for dynamic CBCT perfusion, and second, it saves computation time due to the lower number of reconstructions that need to be performed. Here, only four basis functions are enough to describe the dynamic behaviour of the liver. In addition, these basis functions were only extracted from the CT time-series volumes of two different animals (see Figure 3), with a slightly different contrast agent injection protocol, and were used to model the dynamics of the third animal. This was particularly important to show that the prior-CT scans might not be required for extraction of PK for new subjects. The results show that the TST not only provides the perfusion maps quickly to the radiologist, but also makes the perfusion maps more accurate by compensating for the low temporal sampling. The TST with analytical or PK basis functions can outperform static reconstruction, especially for higher noise levels (see Table 3). The AIF was always best modeled by the TST with PK and the TST with PK resulted in perfusion maps more similar to CT perfusion maps (see Figure 11).

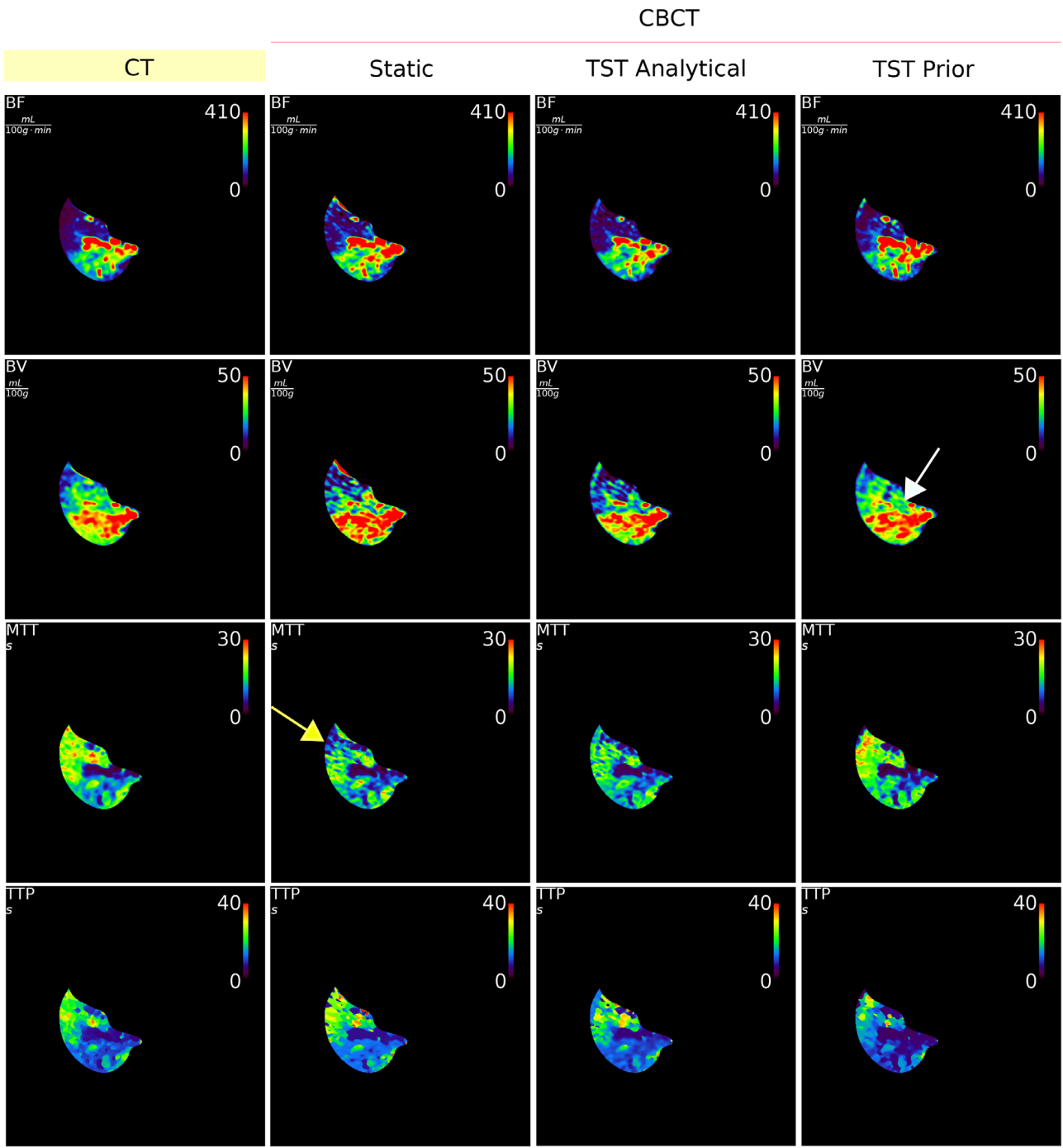

**FIGURE 11** Perfusion maps BF, BV, MTT and TTP for dCTp and reprojected dCBCTp scan (static reconstruction, model-based reconstruction with analytical and PK basis functions) with moderate noise added. The white arrow is pointing to underestimate BV in PK and the yellow arrow is pointing to misestimated region due to reconstruction artifacts in MTT for static reconstruction. BF, blood flow; BV, blood volume; dCTp, dynamic computed tomography perfusion; dCBCTp, cone-beam CT perfusion; MTT, mean transit time; PK, prior knowledge; TTP, time to peak.

The TST with PK is influenced by many factors that cannot be easily addressed. Here, only two CT scans were available for PK extraction, and they could not be averaged enough with SVD to try to avoid the necessity of time alignment of CT scans. The time alignment is performed on the basis of the AIF peak position, which is affected by the anatomy of the scanned subjects, the contrast agent injection protocol, and the actual localization AIF. Next, the scalar product should be calculated, but then insufficient data is available for use to model the dynamics and the interpolation performed based on the very small number of points. Also, depending on the basis functions shift, either the PK will get violated through the reorthogonalization or, the CBCT data will be lost due to the exclusion of the projections for which the PK is not provided. However, with all these demerits, the TST still managed to model the perfusion correctly under high noise.

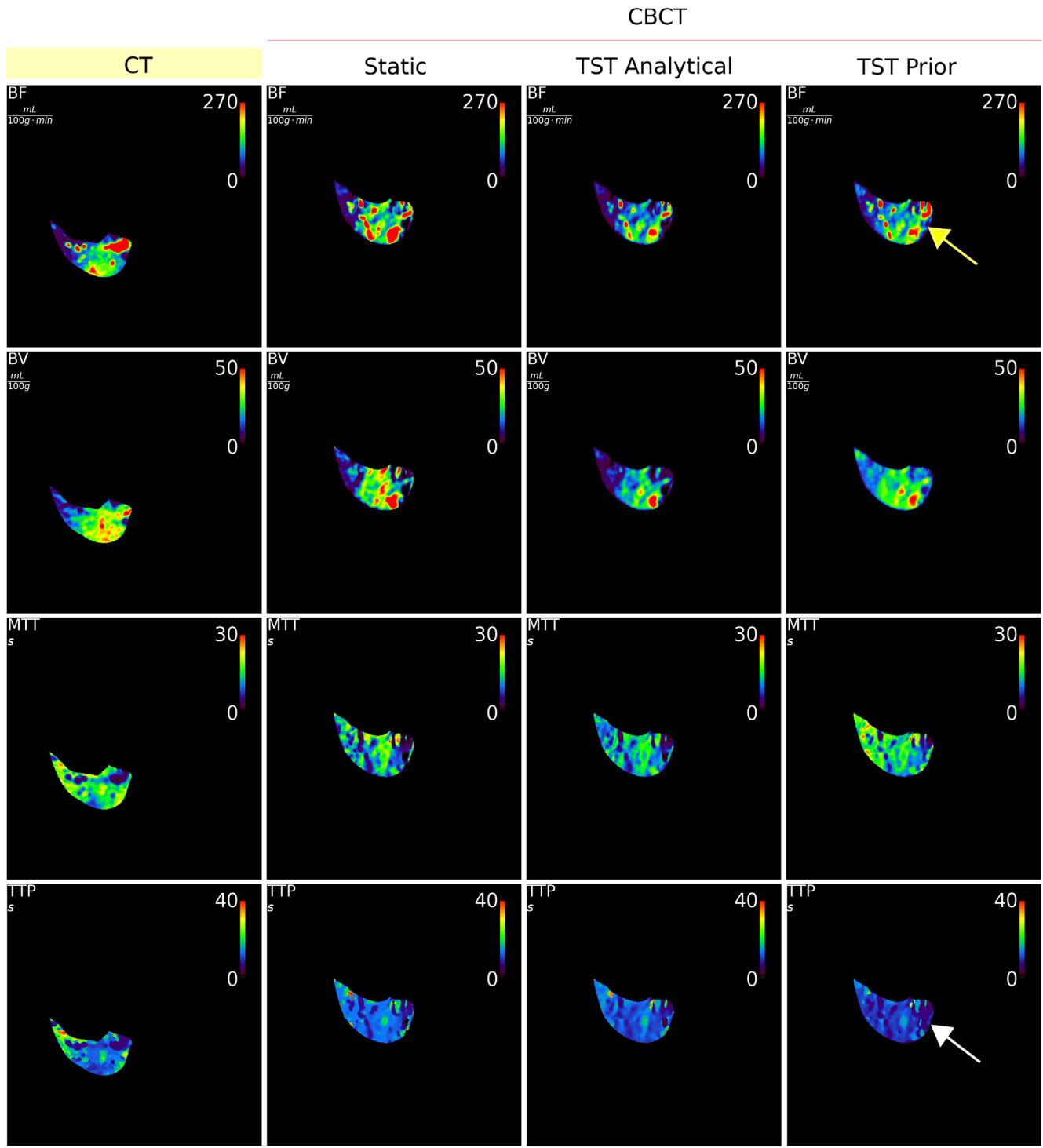

**FIGURE 12** Perfusion maps BF, BV, MTT and TTP calculated using deconvolution method for dCTp and real (i.e., acquired with CBCT) dCBCTp scan (static reconstruction, model-based reconstruction with analytical and PK basis functions). The yellow arrow is pointing to overestimated hypo-perfused region in BF for PK. The white arrow is pointing to misestimated TTP for PK. BF, blood flow; BV, blood volume; CBCT, cone-beam CT; dCTp, dynamic computed tomography perfusion; dCBCTp, cone-beam CT perfusion; MTT, mean transit time; PK, prior knowledge; TTP, time to peak.

CT scans cannot be considered completely noise-free. When the SVD is applied to extract the PK, there is a possibility that some of the noise remains present in the singular vectors forming the BFS. Nevertheless, the noise affecting the accuracy of perfusion maps with TST is reduced; see Table 4. In addition to noise and low temporal sampling, it would be useful to confirm the robustness of TST to scattering. However, such simulations were beyond the scope of the current work. For the extraction of the basis functions, the TACs of CT time-series are interpolated using Akima splines, which can result in error, since with the cubic splines the value between two points can be overestimated.

Although PK extracted from only two animals was sufficient to reconstruct the dCBCTp scan, it is not clear if the same would be possible for any dCBCTp scan, especially with a different contrast agent injection protocol. Next, the AIF with the BFS formed with more than four basis functions, namely seven, would have a higher amplitude.[24] However, using more basis functions causes instabilities in fitting, and inability to produce meaningful perfusion maps.[16,24] Here, already four basis functions were used to model seven points. Therefore, the number of basis functions should be kept as low as possible.

The AIF was selected at one single point in static reconstructions, and this makes it susceptible to higher noise. Also, without the 3D registration of the real CBCT and CT volumes, the selection of the AIF in both is only based on the anatomy knowledge. Thus, the AIF could be calculated by weighted averaging of all the voxels from the selected ROI. Without the static reconstruction, the AIF should be selected in one of the reconstructed TST coefficients. For doing so, one should know which of the basis functions contains the most information about the vessels' dynamics and then select the appropriate reconstructed coefficient to select the AIF. This imposes problems for the automation of the AIF selection within the TST approach. Just by selecting a slightly different AIF location or, mistakenly, a completely different vessel in CBCT compared to CT, might result in a completely wrong estimation of the perfusion maps.[32] However, it is valuable to note that the actual values of perfusion parameters do not differ much between the comparing perfusion maps and that not for all perfusion parameters. Using the deconvolution method for the estimation of perfusion parameters, the chances of misestimation are lower. With the maximum slope, the values and accuracy would primarily depend on the maximum slope of every TAC, which in the case of secondary peaks, such as for the analytical basis functions, and not only for the AIF, would contribute to the poor assessment of the perfusion parameters. Alternatively, gamma-variate functions could be fitted to the AIF[35] and to all the TACs, to smooth them out and leave out noise. Not only should the large number of TACs be fitted, but also for each one, the fitting parameters should be adjusted. This is particularly difficult because of the noise that results in multiple peaks in the TAC. At first glance, the idea of fitting a gamma variate to also hypo-perfused areas might seem good, but these regions are also affected by noise, and even though the peaks in such fitted TACs would be expected to be very low, they could still be wrong in the time domain and falsely represent perfusion.

The contrast agent flow is highly dependent on the anatomy of the subject being scanned. This means that even for exactly the same contrast agent injection protocol, the basis functions might need to be shifted in time. In this work, this was simulated in such a way that the rdCBCTp had to be shifted in time. The quality of the perfusion maps did not suffer from it even though all the projections acquired in the first 1.5 s were excluded, and for these views only seven points could then be used for modeling. Within the real CBCT scan, it was sufficient to simply exclude the mask sweeps. However, by excluding mask sweeps, the baseline needed for the subtraction of static anatomical structures is lost, so the values presented in the perfusion maps are not true to nature. Considering the outflow of the contrast agent and little to no presence in rest of the sweeps, instead of the last two sweeps, two mask sweeps at the beginning could be acquired, without acquiring more than eight sweeps. The time shift could already be estimated in the projection domain by fitting the basis functions to regions consisting of mostly vessels and selecting the shift with lowest fitting error, which would be very time-consuming. Alternatively, this could be done if the position of the AIF could be assumed in projection domain, since the vessels are visible in contrast-enhanced projections.

The TST is not robust to motion and the motion correction step within it could only be performed in projection domain before the fitting of the basis functions functions. For shallow breathing,[36] the displacement vector could be formed based on the diaphragm position.[37] Otherwise, the time-resolved volumes would need to be calculated and registered. The number of points used for TAC sampling for perfusion parameter calculation determines the number of volumes for registration. Furthermore, the rigid registration is not sufficient to compensate for the patient motion, breathing motion, and as a consequence of the breathing, the motion inside the liver. Although breathing motion was not present in the data from this study, motion correction would still be beneficial for motion in the intestines and for the misalignment between the acquired views, due to the inconsistency between the corresponding angles in forward and backward sweep.

CT and CBCT imaging protocols differ in acquisition speed, detector response, scattering, dynamic range, and spatial resolution, all of which can significantly impact perfusion measurements.[3] To address these challenges, a model-based approach (TST) was utilized, which is specifically designed to mitigate the effects of low temporal sampling and noise in CBCT imaging. Both simulated CBCT data—based on reprojected CT images—and real CBCT perfusion scans, acquired using animal models, were included to account for differences between CT and CBCT and validate the robustness of our approach. This combined methodology enables us to overcome practical limitations while demonstrating the feasibility of CBCT perfusion imaging for hepatic applications. Further analysis on how to optimize the dose and protocol is needed before possible clinical applications of the method, however this is out of scope of the present paper.

## 5 | CONCLUSION AND FUTURE WORK

This research explored the potential of CBCT for dynamic liver perfusion imaging, focusing on the value of MBR using the TST. Although a marked improvement over standard reconstruction methods was not universally observed, our study demonstrated that TST, with a set of basis functions derived from only two dynamic CT perfusion scans from different animals, can outperform static reconstruction in a simulated dynamic CBCT perfusion scan. This suggests that a prior CT scan of the patient may not always be necessary.

Furthermore, this study is the first to demonstrate that MBR with TST using PK can achieve accurate results in real dynamic CBCT perfusion scans of the liver. However, confirmation using larger datasets is needed to establish statistical robustness, along with a detailed comparison of CBCT and CT perfusion maps, including the registration of CBCT and CT volumes. Additionally, the influence of varying contrast agent injection protocols warrants further investigation, and exploring how anatomical knowledge of the organ can enhance basis extraction is another promising direction for future work.

This study reaffirms the broader value of TST as a flexible technique with potential applications beyond perfusion imaging. Methods based on fitting a set of basis functions to projection data offer promise across various areas of time-dependent tomography, including noise reduction, motion compensation, and the tracking of CT volumes undergoing motion or developmental changes during scanning. Further research is needed to rigorously investigate its possibilities for various dynamic imaging applications.

## ACKNOWLEDGMENTS
The work of this paper is funded by the German Ministry of Education and Research (BMBF) within the Forschungscampus STIMULATE under grant no. 13GW0473A and 13GW0473B.

Open access funding enabled and organized by Projekt DEAL.

## CONFLICT OF INTEREST STATEMENT
The authors declare no conflicts of interest.

## REFERENCES

1. Ronot M, Lambert S, Daire JL, et al. Can we justify not doing liver perfusion imaging in 2013?. *Diagn Interventional Imaging*. 2013;94(12):1323-1336. doi:10.1016/j.diii.2013.06.005
2. Ogul H, Kantarci M, Genc B, et al. Perfusion CT imaging of the liver: review of clinical applications. *Diagn Interv Radiol*. 2014;20(5):379-389. doi:10.5152/dir.2014.13396
3. Orth RC, Wallace MJ, Kuo MD, et al. C-arm cone-beam CT: general principles and technical considerations for use in interventional radiology. *J Vasc Interv Radiol*. 2008;19(6):814-820. doi:10.1016/j.jvir.2008.02.002
4. Huppert PE, Firlbeck G, Meissner OA, Wietholtz H. C-Arm-CT bei der Chemoembolisation von Lebertumoren. *Der Radiologe*. 2009;49(9):830-836. doi:10.1007/s00117-009-1862-7
5. Peynircioglu B, Hizal M, Cil B, et al. Quantitative liver tumor blood volume measurements by a C-arm CT post-processing software before and after hepatic arterial embolization therapy: comparison with MDCT perfusion. *Diagn Interv Radiol*. 2015;21(1):71-77. doi:10.5152/dir.2014.13290
6. Vogl TJ, Schaefer P, Lehnert T, et al. Intraprocedural blood volume measurement using C-arm CT as a predictor for treatment response of malignant liver tumours undergoing repetitive transarterial chemoembolization (TACE). *Eur Radiol*. 2015;26(3):755-763. doi:10.1007/s00330-015-3869-y
7. Rathmann N, Kara K, Budjan J, et al. Parenchymal liver blood volume and dynamic volume perfusion CT measurements of hepatocellular carcinoma in patients undergoing transarterial chemoembolization. *Anticancer Res*. 2017;37(10):5681-5685. doi:10.21873/anticanres.12004
8. Choi SY, Kim KA, Choi W, Kwon Y, Cho SB. Usefulness of cone-beam CT-Based liver perfusion mapping for evaluating the response of hepatocellular carcinoma to conventional transarterial chemoembolization. *J Clin Med*. 2021;10(4):713. doi:10.3390/jcm10040713
9. Zaid Al-Kaylani AHA, Schuurmann RCL, Maathuis WD, Slart RHJA, Vries JPPM, Bokkers RPH. Clinical applications of quantitative perfusion imaging with a C-arm flat-panel detector— a systematic review. *Diagnostics*. 2022;13(1):128. doi:10.3390/diagnostics13010128
10. Datta S, Müller K, Moore Te, et al. Dynamic measurement of arterial liver perfusion with an interventional C-Arm system. *Invest Radiol*. 2017;52(8):456-461. doi:10.1097/rli.0000000000000368
11. Fieselmann A, Manhart M. C-arm CT Perfusion imaging in the interventional suite. *Curr Med Imaging Rev*. 2013;9(2):96-101. doi:10.2174/1573405611309020004
12. Ortega-Gutierrez S, Quispe-Orozco D, Schafer S, et al. Angiography suite cone-beam CT perfusion for selection of thrombectomy patients: a pilot study. *J Neuroimaging*. 2022;32:493-501. doi:10.1111/jon.12988
13. Zaid Al-Kaylani AHA, Schuurmann RCL, Maathuis WD, Slart RHJA, Vries JPPM, Bokkers RPH. Clinical applications of conebeam CTP imaging in cerebral disease: a systematic review. *AJNR Am J Neuroradiol*. 2023;44(8):922-927. doi:10.3174/ajnr.a7930
14. Kim KA, Choi SY, Kim MU, et al. The efficacy of cone-beam CT–Based liver perfusion mapping to predict initial response of hepatocellular carcinoma to transarterial chemoembolization. *J Vasc Interv Radiol*. 2019;30(3):358-369. doi:10.1016/j.jvir.2018.10.002
15. Neukirchen C, Rose G. Parameter estimation in a model based approach for tomographic perfusion measurement. *IEEE Nucl Sci Symp Conf Rec*. 2005;4:2235-2239. doi:10.1109/nssmic.2005.1596779

16. Bannasch S, Frysch R, Pfeiffer T, Warnecke G, Rose G. Time separation technique: Accurate solution for 4D C-Arm-CT perfusion imaging using a temporal decomposition model. *Med Phys*. 2018;45:1080-1092. doi:10.1002/mp.12768
17. Kulvait V, Hoelter P, Frysch R, Haseljić H, Doerfler A, Rose G. A novel use of time separation technique to improve flat detector CT perfusion imaging in stroke patients. *Med Phys*. 2022;49(6):3624-3637. doi:10.1002/mp.15640
18. Haseljić H, Kulvait V, Frysch R, et al. The application of time separation technique to enhance C-arm CT dynamic liver perfusion imaging. In: *Proceedings of the 16th Virtual International Meeting on Fully 3D Image Reconstruction in Radiology and Nuclear Medicine*. arxiv; 2021. doi:10.48550/ARXIV.2110.04143
19. Akima Hiroshi. A New Method of Interpolation and Smooth Curve Fitting Based on Local Procedures. *J ACM*. 1970;17(4):589-602. doi:10.1145/321607.321609
20. Pfeiffer T, Frysch R, Bismark RN, Rose G. CTL: modular open-source C++-library for CT-simulations. In: *15th International Meeting on Fully Three-Dimensional Image Reconstruction in Radiology and Nuclear Medicine*. SPIE; 2019:38. doi:10.1117/12.2534517
21. Manhart MT, Kowarschik M, Fieselmann A, et al. dynamic iterative reconstruction for interventional 4-D C-Arm CT perfusion imaging. *IEEE Trans Med Imaging*. 2013;32(7):1336-1348. doi:10.1109/tmi.2013.2257178
22. Manhart MT, Aichert A, Struffert T, et al. Denoising and artefact reduction in dynamic flat detector CT perfusion imaging using high speed acquisition: first experimental and clinical results. *Phys Med Biol*. 2014;59(16):4505-4524. doi:10.1088/0031-9155/59/16/4505
23. Kulvait V, Rose G. Software Implementation of the Krylov Methods Based Reconstruction for the 3D Cone Beam CT Operator. In: *16th International Meeting on Fully Three-Dimensional Image Reconstruction in Radiology and Nuclear Medicine*. arXiv; 2021. doi:10.48550/arXiv.2110.13526
24. Haseljić H, Kulvait V, Frysch R, et al. Time separation technique using prior knowledge for dynamic liver perfusion imaging. In: Stayman JW, ed. *7th International Conference on Image Formation in X-Ray Computed Tomography*. SPIE; 2022:32. doi:10.1117/12.2646449
25. Eckel C, Bannasch S, Frysch R, Rose G. A compact and accurate set of basis functions for model-based reconstructions. *Curr Dir Biomed Eng*. 2018;4(1):323-326. doi:10.1515/cdbme-2018-0078
26. Falini A. A review on the selection criteria for the truncated SVD in Data Science applications. *JCMDS*. 2022;5:100064. doi:10.1016/j.jcmds.2022.100064
27. Fieselmann A, Ganguly A, Deuerling-Zheng Y, et al. Interventional 4-D C-Arm CT perfusion imaging using interleaved scanning and partial reconstruction interpolation. *IEEE Trans Med Imaging*. 2012;31(4):892-906. doi:10.1109/tmi.2011.2181531
28. Haseljic H, Frysch R, Werncke T, Rose G. Comparison of Deconvolution and Maximum Slope Method in Dynamic CBCT Liver Perfusion Imaging for Evaluation of Performed Embolization. *6th IGIC conference*. 2023. Accessed May 4, 2024. https://www.igic.de/deutsch/igic-2023/proceedings
29. Kanda T, Yoshikawa T, Ohno Y, et al. Hepatic computed tomography perfusion: comparison of maximum slope and dual-input single-compartment methods. *Jpn J Radiol*. 2010;28(10):714-719. doi:10.1007/s11604-010-0497-y
30. Fieselmann A, Kowarschik M, Ganguly A, Hornegger J, Fahrig R. Deconvolution-based CT and MR brain perfusion measurement: theoretical model revisited and practical implementation details. *Int J Biomed Imaging*. 2011;2011:467563. doi:10.1155/2011/467563
31. Kudo K. *Acute Stroke Imaging Standardization Group Japan recommended standard LUT (a-LUT) for perfusion color maps*. 2005. Accessed April 9, 2024. http://asist.umin.jp/data-e.shtml
32. Haseljic H, Frysch R, Kulvait V, Werncke T, Speck O, Rose G. The impact of arterial input function selection on dynamic liver perfusion imaging with cone-beam CT. *CT Meeting 2024*. 2024. Accessed November 19, 2024. https://www.ct-meeting.org/data/ProceedingsCTMeeting2024.pdf
33. Donoho DL, Johnstone IM. Ideal spatial adaptation by wavelet shrinkage. *Biometrika*. 1994;81(3):425-455. doi:10.1093/biomet/81.3.425
34. Kulvait V, Hoelter P, Punzet D, Doerfler A, Rose G. Noise and dose reduction in CT brain perfusion acquisition by projecting time attenuation curves onto lower dimensional spaces. Proc. SPIE 12031. L. *Medical Imaging 2022: Physics of Medical Imaging*. 2022, 1203133. doi:10.1117/12.2611432
35. D'Anto M, Cesarelli M, Bifulco P, et al. Study of different Time Attenuation Curve processing in liver CT perfusion. *Proceedings of the 10th IEEE International Conference on Information Technology and Applications in Biomedicine*. IEEE; 2010:1-4. doi:10.1109/ITAB.2010.5687750
36. Tacher V, Radaelli A, Lin M, Geschwind JF. How i do it: cone-beam ct during transarterial chemoembolization for liver cancer. *Radiology*. 2015;274(2):320-334. doi:10.1148/radiol.14131925
37. Rit S, Wolthaus JWH, Herk M, Sonke JJ. On-the-fly motion-compensated cone-beam CT using an a priori model of the respiratory motion. *Med Phys*. 2009;36(6Part1):2283-2296. doi:10.1118/1.3115691
38. Kaplan D. Knee Point. MATLAB Central File Exchange. 2024. Accessed May 15, 2024. https://www.mathworks.com/matlabcentral/fileexchange/35094-knee-point
39. Satopaa V, Albrecht J, Irwin D, Raghavan B. Finding a "Kneedle" in a haystack: detecting knee points in system behavior. In: *2011 31st International Conference on Distributed Computing Systems Workshops*. IEEE; 2011. doi:10.1109/icdcsw.2011.20

## APPENDIX A: PK BFS

The number of basis functions which would form the PK BFS was selected so it models the perfusion correctly while keeping this number the lowest possible. The two criteria are interconnected. More basis functions could indeed potentially describe the perfusion better, but at the same time, it could bring instabilities to the system of linear equations for estimation of fitting coefficients. The CBCT scan consists of eight sweeps. However, due to time shifting of the basis function set, it can be that not all eight points are included into fitting. Therefore, with more than seven or eight basis functions the system would be overdetermined. With eight basis functions, the number of reconstructions would already equal the total number of static reconstructions. Moreover, it is unlikely that eight basis functions could improve the overall quality of perfusion maps compared to those calculated out of static reconstructions.

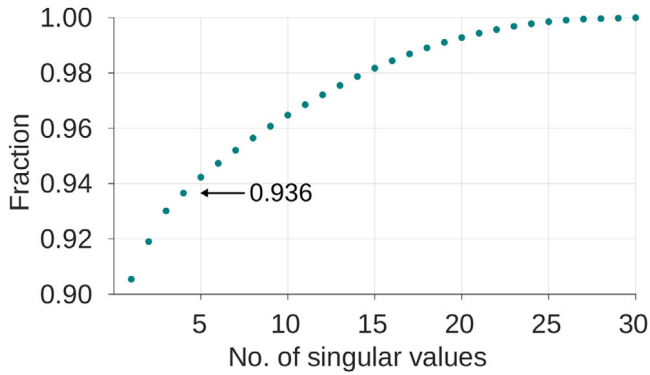

**FIGURE A1** Sum of *n* first singular values divided by the sum of the first 30 singular values. The arrow is pointing to the point which determines the basis of four functions.

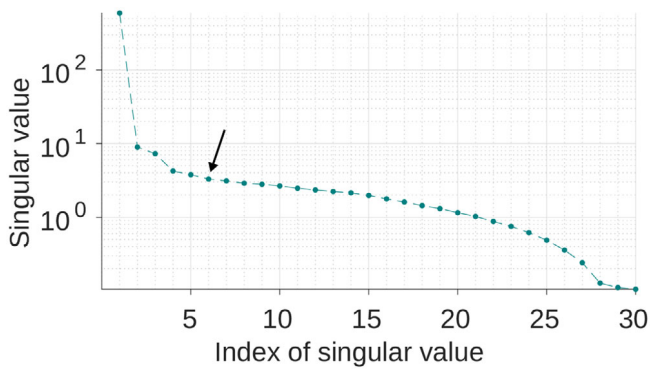

**FIGURE A2** First 30 singular values on logarithmic scale. The arrow is pointing to an elbow point.

One way of determining the number of basis functions could be by defining the fraction to which the sum of *n* singular values divided by the sum of all 30 singular values is equal or is the closest smaller. The curve is converging to 1. For $n = 4$ singular values, this fraction would be

$$\frac{\sum_{i=1}^{4} s_i}{\sum_{i=1}^{30} s_i} = 0.936. \quad (A1)$$

**TABLE A1** Pearson correlation coefficient for rCBCT perfusion maps for different number of basis functions compared to CT perfusion maps. The highet values are highlighted in bold.

| | Number of basis functions forming the set | | | | | | |
|---|---|---|---|---|---|---|---|
| | 2 | 3 | 4 | 5 | 6 | 7 | 8 |
| BF | 0.7374 | 0.9126 | **0.9371** | 0.9291 | 0.9305 | 0.9057 | 0.9029 |
| BV | 0.7134 | 0.8122 | **0.8475** | 0.7867 | 0.7897 | 0.6670 | 0.5696 |
| MTT | 0.6516 | 0.8162 | **0.8220** | 0.7461 | 0.7527 | 0.5448 | 0.6544 |
| TTP | 0.7450 | 0.7563 | **0.7620** | 0.6645 | 0.6769 | 0.5025 | 0.5737 |

Abbreviations: BF, blood flow; BV, blood volume; MTT, mean transit time; TST, time separation technique; TTP, time to peak; TST, time separation technique.

In Figure A1 the sum can be observed for the first 30 singular values. To obtain larger than 0.95, seven vectors would need to form the BFS, which would result in only one less reconstruction than for static reconstruction approach.

The number of basis functions has also been determined by the elbow method.[38] For this method, the selection of the first *n* singular values for singular values function, for which the elbow will be estimated, influences the outcome as well.[39] Therefore, the first singular value was left out, considering that $s1 = 594.03$ and $s2 = 8.94$. It is expected to have a high value of the first singular value considering the almost constant first singular vector which indicates it represents the static behaviour of the organ. The elbow indicates that the basis function set should be formed from the first six singular vectors, see Figure A2.

However, to rely solely on these assessments is not a guarantee that the appropriate basis function set has been found. The elbow method has been a good indicator where to start lowering the number of basis functions forming the set, and this was to be analyszd by calculating the Pearson correlation for perfusion maps, see Table A1. The largest correlation coefficients are obtained for the basis function set consisting of four basis functions.